\newcommand\apjcls{1}
\newcommand\aastexcls{2}
\newcommand\othercls{3}

\newcommand\papercls{\aastexcls}
\documentclass[tighten, times,twocolumn]{aastex62}  % ApJ 

\if\papercls \apjcls
\usepackage{apjfonts}
\else\if\papercls \othercls
\usepackage{epsfig}
\usepackage{margin}
\usepackage{times}
\fi\fi
\usepackage{ifthen}
\usepackage{natbib}
\usepackage{bm}
\usepackage{amssymb, amsmath}
\usepackage{appendix}
\usepackage{etoolbox}
\usepackage[T1]{fontenc}
\usepackage{paralist}
\usepackage{newtxtext,newtxmath}
\if\papercls \apjcls 
\newcommand\aas{\ref@jnl{AAS Meeting Abstracts}}% *** added by jh
\newcommand\dps{\ref@jnl{AAS/DPS Meeting Abstracts}}% *** added by jh
\newcommand\maps{\ref@jnl{MAPS}}% *** added by jh
\else\if\papercls \othercls
\usepackage{astjnlabbrev-jh}
\fi\fi

\if\papercls \aastexcls
\hypersetup{citecolor=blue, % color for \cite{...} links
            linkcolor=blue, % color for \ref{...} links
            menucolor=blue, % color for Acrobat menu buttons
            urlcolor=blue}  % color for \url{...} links
\else
\usepackage[%pdftex,      % hyper-references for pdflatex
bookmarks=true,           %%% generate bookmarks ...
bookmarksnumbered=true,   %%% ... with numbers
colorlinks=true,          % links are colored
citecolor=blue,           % color for \cite{...} links
linkcolor=blue,           % color for \ref{...} links
menucolor=blue,           % color for Acrobat menu buttons
urlcolor=blue,            % color for \url{...} links
linkbordercolor={0 0 1},  %%% blue frames around links
pdfborder={0 0 1},
frenchlinks=true]{hyperref}
\fi

\if\papercls \othercls

\else

\fi

\providecommand{\adsurl}[1]{\href{#1}{ADS}}

\makeatletter
\patchcmd{\NAT@citex}
  {\@citea\NAT@hyper@{%
     \NAT@nmfmt{\NAT@nm}%
     \hyper@natlinkbreak{\NAT@aysep\NAT@spacechar}{\@citeb\@extra@b@citeb}%
     \NAT@date}}
  {\@citea\NAT@nmfmt{\NAT@nm}%
   \NAT@aysep\NAT@spacechar\NAT@hyper@{\NAT@date}}{}{}

\patchcmd{\NAT@citex}
  {\@citea\NAT@hyper@{%
     \NAT@nmfmt{\NAT@nm}%
     \hyper@natlinkbreak{\NAT@spacechar\NAT@@open\if*#1*\else#1\NAT@spacechar\fi}%
       {\@citeb\@extra@b@citeb}%
     \NAT@date}}
  {\@citea\NAT@nmfmt{\NAT@nm}%
   \NAT@spacechar\NAT@@open\if*#1*\else#1\NAT@spacechar\fi\NAT@hyper@{\NAT@date}}
  {}{}
\makeatother

\makeatletter
\DeclareRobustCommand{\lowcase}[1]{\@lowcase#1\@nil}
\def\@lowcase#1\@nil{\if\relax#1\relax\else\MakeLowercase{#1}\fi}
\makeatother

\DeclareSymbolFont{UPM}{U}{eur}{m}{n}
\DeclareMathSymbol{\umu}{0}{UPM}{"16}
\let\oldumu=\umu
\renewcommand\umu{\ifmmode\oldumu\else\math{\oldumu}\fi}

\if\papercls \othercls

\else

\fi

\let\oldsim=\sim
\renewcommand\sim{\ifmmode\oldsim\else\math{\oldsim}\fi}
\let\oldpm=\pm
\renewcommand\pm{\ifmmode\oldpm\else\math{\oldpm}\fi}
\newcommand\by{\ifmmode\times\else\math{\times}\fi}
\newbox{\wdbox}
\renewcommand\c{\setbox\wdbox=\hbox{,}\hspace{\wd\wdbox}}
\renewcommand\i{\setbox\wdbox=\hbox{i}\hspace{\wd\wdbox}}

\newcount\timect
\newcount\hourct
\newcount\minct
\newcommand\now{\timect=\time \divide\timect by 60
         \hourct=\timect Cltiply\hourct by 60
         \minct=\time \advance\minct by -\hourct
         \number\timect:\ifnum \minct < 10 0\fi\number\minct}

\catcode`@=11

\newcommand\comment[1]{}

\newcommand\commenton{\catcode`\%=14}

\renewcommand\math[1]{$#1$}
\newcommand\mathshifton{\catcode`\$=3}

\let\atab=&
\newcommand\atabon{\catcode`\&=4}

\let\oldmsp=\sp
\let\oldmsb=\sb
\def\sp#1{\ifmmode
           \oldmsp{#1}%
         \else\strut\raise.85ex\hbox{\scriptsize #1}\fi}
\def\sb#1{\ifmmode
           \oldmsb{#1}%
         \else\strut\raise-.54ex\hbox{\scriptsize #1}\fi}
\newbox\@sp
\newbox\@sb
\def\sbp#1#2{\ifmmode%
           \oldmsb{#1}\oldmsp{#2}%
         \else
           \setbox\@sb=\hbox{\sb{#1}}%
           \setbox\@sp=\hbox{\sp{#2}}%
           \rlap{\copy\@sb}\copy\@sp
           \ifdim \wd\@sb >\wd\@sp
             \hskip -\wd\@sp \hskip \wd\@sb
           \fi
        \fi}
\def\msp#1{\ifmmode
           \oldmsp{#1}
         \else \math{\oldmsp{#1}}\fi}
\def\msb#1{\ifmmode
           \oldmsb{#1}
         \else \math{\oldmsb{#1}}\fi}

\def\supon{\catcode`\^=7}

\def\subon{\catcode`\_=8}

\def\supsubon{\supon \subon}

\newcommand\actcharon{\catcode`\~=13}

\newcommand\paramon{\catcode`\#=6}

\comment{And now to turn us totally on and off...}

\newcommand\reservedcharson{ \commenton  \mathshifton  \atabon  \supsubon 
                             \actcharon  \paramon}

\catcode`@=12
\reservedcharson

\if\papercls \apjcls

\else

\fi

\newcommand\chisq{\ifmmode{\chi\sp{2}}\else\math{\chi\sp{2}}\fi}
\newcommand\redchisq{\ifmmode{ \chi\sp{2}\sb{\rm red}}
                    \else\math{\chi\sp{2}\sb{\rm red}}\fi}
\newcommand\Teq{\ifmmode{T\sb{\rm eq}}\else$T$\sb{eq}\fi}
\newcommand\mjup{\ifmmode{M\sb{\rm Jup}}\else$M$\sb{Jup}\fi}
\newcommand\rjup{\ifmmode{R\sb{\rm Jup}}\else$R$\sb{Jup}\fi}
\newcommand\msun{\ifmmode{M\sb{\odot}}\else$M\sb{\odot}$\fi}
\newcommand\rsun{\ifmmode{R\sb{\odot}}\else$R\sb{\odot}$\fi}
\newcommand\mearth{\ifmmode{M\sb{\oplus}}\else$M\sb{\oplus}$\fi}
\newcommand\rearth{\ifmmode{R\sb{\oplus}}\else$R\sb{\oplus}$\fi}
\renewcommand{\bm}[1]{{\mbox{{\boldmath$#1$}}}}	% bold in mathmode

\newcommand{\grad}{\bm{\nabla}}
\newcommand{\cross}{\bm{\times}}

\shorttitle{Assessing the Observability of Deep Meridional Flow Cells in the Solar Interior}
\shortauthors{Fuentes {\em et al.}}

\begin{document}

%\title{Systematic Bias in Helioseismic Measurements of Meridional Circulation Arising from Nonlocal Averaging Kernels}
%\title{Unveiling the Solar Meridional Circulation: Insights from Ray-theory Averaging Kernels and Hydrodynamical Simulations}
\title{Assessing the Observability of Deep Meridional Flow Cells in the Solar Interior}

\author{J. R. Fuentes}
\affiliation{\rm Department of Applied Mathematics, University of Colorado Boulder, Boulder, CO 80309-0526, USA}

\author{Bradley W. Hindman}
\affiliation{\rm Department of Applied Mathematics, University of Colorado Boulder, Boulder, CO 80309-0526, USA}
\affiliation{\rm  JILA \& Department of Astrophysical and Planetary Sciences, University of Colorado Boulder, Boulder, CO 80309-0440, USA}

\author{Junwei Zhao}
\affiliation{\rm W. W. Hansen Experimental Physics Laboratory, Stanford University, Stanford, CA 94305-4085, USA
}
\author{Catherine C. Blume}
\affiliation{\rm  JILA \& Department of Astrophysical and Planetary Sciences, University of Colorado Boulder, Boulder, CO 80309-0440, USA}

\author{Maria E. Camisassa}
\affiliation{\rm Departament de F\'\i sica, Universitat Polit\`ecnica de Catalunya, c/Esteve Terrades 5, 08860 Castelldefels, Spain}

\author{Nicholas A. Featherstone}
\affiliation{\rm Southwest Research Institute, Department of Solar and Heliospheric Physics, Boulder, CO 80302, USA
}

\author{Thomas Hartlep}
\affiliation{\rm Bay Area Environmental Research Institute, Moffett Field, CA 94035, USA
}

\author{Lydia Korre}
\affiliation{\rm Department of Applied Mathematics, University of Colorado Boulder, Boulder, CO 80309-0526, USA}

\author{Loren I. Matilsky}
\affiliation{\rm Department of Applied Mathematics, Baskin School of Engineering, University of California, Santa Cruz, CA 95064, USA
}

\begin{abstract}

Meridional circulation regulates the Sun's interior dynamics and magnetism. While it is well accepted that meridional flows are poleward at the Sun's surface, helioseismic observations have yet to provide a definitive answer for the depth at which those flows return to the equator, or the number of circulation cells in depth. Here, we explore the observability of multiple circulation cells stacked in radius. Specifically, we examine the seismic signature of several meridional flow profiles by convolving time-distance averaging kernels with mean flows obtained from a suite of 3D hydrodynamic simulations. At mid and high latitudes, we find that weak flow structures in the deep convection zone can be obscured by signals from the much stronger surface flows. This contamination of 1--2 m s$^{-1}$ is caused by extended side lobes in the averaging kernels, which produce a spurious equatorward signal with flow speeds that are one order of magnitude stronger than the original flow speeds in the simulations. At low latitudes, the flows in the deep layers of the simulations are stronger ($> 2$ m s$^{-1}$) and multiple cells across the convection zone can produce a sufficiently strong signal to survive the convolution process. Now that meridional flows can be measured over two decades of data, the uncertainties arising from convective noise have fallen to a level where they are comparable in magnitude to the systematic biases caused by non-local features in the averaging kernels. Hence, these systematic errors are beginning to influence current helioseismic deductions and need broader consideration.

\end{abstract}


\section{Introduction} 

The meridional circulation of the Sun is an axisymmetric flow pattern in which material moves from the equator towards the poles at the surface.  Because of mass conservation, the flow submerges into the interior, switches direction, and returns towards the equator at some depth. This flow is of great relevance to the global dynamics of the Sun and to its magnetism because it may play an important role in modulating the amplitude and duration of sunspot cycles \citep[e.g.,][]{Jiang_2010,Upton_2014b,Upton_2014a} and at transporting angular momentum and magnetic flux \citep[e.g.,][]{Wang_89,Miesch_2005,Featherstone_2015}. Yet, characterizing the structure of the meridional circulation has been a challenge. %due to its relatively low amplitude compared to the background convective turbulence and due to systematic errors in the observational procedures, such as the well-known but poorly understood center-to-limb effect \citep[e.g.,][]{Zhao_2012}}.

Over the last three decades, many observational attempts have been made to measure the solar meridional circulation  \citep[see,][for a recent review]{Hanasoge_2022}. There is a general concensus that at the Sun’s surface and in a shallow layer below (i.e., from the photosphere to a depth of $\sim30$ Mm), the meridional flow is predominantly poleward, with a speed of $10$--$20 ~\mathrm{m~s^{-1}}$ \citep[e.g.,][]{Komm_1993,Svanda_2007,Ulrich_2010,Zhao_2012, Hathaway_2012,Hathaway2022}. Nowadays, improved helioseismic techniques have enabled measurements down to 200 Mm in depth, i.e., near the bottom of the convection zone. However, these extended measurements are inconclusive; different helioseismic analyses of the same data and the same analysis applied to data from different instruments produce different results. In particular, there is evidence for both a single-cell meridional circulation with a deep return flow \citep[e.g.,][]{Giles_2000,Rajaguru_2015,Gizon_2020} and a double-cell meridional circulation with a return flow near the middle of the convection zone \citep[e.g.,][]{Hathaway_2012,Zhao_2013,Chen_and_Zhao_2017, Jackiewicz_et_al_2015, Lin_Chou_2018}. Interestingly, there is even evidence for both morphologies within the same study. For example, a careful examination of the results of \citet{Gizon_2020}, suggests single cells during magnetically active time periods and multiple cells during solar minimum (see their Figure 2c).

There are many potential reasons for such inconsistency, some arising from the travel-time measurements and others from the inversion procedure. A prominent issue concerning the measurements is the removal of systematic biases, particularly the so-called center-to-limb effect \citep{Zhao_2012,Zhao_et_al_2016,Chen_and_Zhao_2018}. Since the physical origin of this bias is unclear, a variety of methods have been suggested for its removal. Complicating this issue is that the systematic biases depend on the mode frequency, radial order, instrumentation, and spectral line choice \citep[e.g.][]{Zhao_2012, Greer_2013, Chen_and_Zhao_2017, Chen_and_Zhao_2018, Rajaguru_2020}.

Many inversion procedures with differing physical assumptions have been developed to measure meridional circulations, and many of the inconsistencies noted previously are likely the result of these differences. For example, some inversion procedures impose mass conservation as a physical constraint on the meridional flow \citep{Schad2013, Rajaguru_2015, Mandal_2017, Gizon_2020}, and others do not \citep{Zhao_2013, Chen_and_Zhao_2017}. The application of such a constraint does seem to typically result in a single flow cell with depth.  However, since direct seismic imaging becomes relatively weak below 0.86 $R_\odot$, the deep flows in such inversions are largely imposed by an interplay of the constraint and the nature of the inversion's regularization. Since there is a continuum of flow structures that conserve mass, it is unclear whether multiple cells with weak flow could be lurking near the bottom of the convection zone.

In addition to potential assumptions about the flow structure, helioseismic inversions also make assumptions about the nature of acoustic wave propagation. Some procedures employ kernels computed using ray theory, and others use the Born Approximation. The former assumes that the Sun’s acoustic waves have infinitely small wavelengths, whereas the latter considers the finite wavelength and scattering of acoustic waves. While, it is indeed true that the averaging kernels produced by the two approximations are different \citep[e.g.][]{Mandal2018}, the two techniques often produce similar results \citep[e.g., see][who used the Born approximation and found that both single and multiple-cell meridional flow profiles can be consistent with their measurements]{Boning2017}.

Finally, another possibility that has received less attention is that the spatial averaging that is inherent in helioseismic measurements could mask the weaker flows of the meridional circulation in the deep interior.  Specifically, the well-known presence of spatial side lobes in the averaging kernels can result in contamination of deep flows with signals from the flows near the surface. In this work, we investigate this possibility by seeking the answer to the following question: If the Sun were to have a meridional circulation similar to those achieved in the numerical simulations, how would the helioseismically observed flow appear?
Previous authors have investigated this problem by generating synthetic wavefield data with various meridional-circulation models, and then performed helioseismic measurements to see whether the pre-set flow profiles can be recovered by helioseismic measurements \citep{Hartlep_2013, Stejko_2023}.  This work uses a complementary approach.

We generate synthetic observations by convolving averaging kernels obtained from the ray approximation \citep[based on those of][]{Zhao_2013} with mean flows obtained from a suite of state-of-the-art numerical simulations of convection in the Sun. We chose this particular set of averaging kernels because they are easily accessible and represent a broad class of currently used inversion techniques. We acknowledge that ray-approximation kernels have limitations \citep[e.g.,][]{Birch2001,Birch2004}; however, the existence of single versus multiple meridional circulation cells does not depend solely on the choice of ray theory \citep[e.g.,][]{Rajaguru_2015, Boning2017}. Thus, the averaging kernels presented in \cite{Zhao_2013} are suitable for exploring the issues we are interested in, such as "side-lobe contamination" and whether or not the deep cell structures that appear in our simulations can be recovered after the convolution.

This paper is organized as follows. In Sections~\ref{sec:sim} and \ref{sec:kernels}, we describe the meridional flows computed from the numerical simulations and the averaging kernels that we apply to them, respectively. Then, in Section~\ref{sec:analysis} we present and analyze the synthetic observations. Finally, in Section~\ref{sec:disc} we summarize our results and discuss their implications for our understanding of the deep solar meridional circulation.

\section{Numerical simulations}  \label{sec:sim}

In this section, we provide a brief overview of our numerical simulations and the resulting flows that we will analyze further in the paper.

\subsection{Fluid equations}

The simulations used in this work correspond to models of the solar convection zone in a 3D spherical shell, including the effects of rotation and density stratification. The simulations were conducted using the open-source \textit{Rayleigh} convection code \citep{Rayleigh_code_2021}, which applies the pseudo-spectral techniques described in \cite{Glatzmaier_1984}, to solve the anelastic fluid equations \citep[e.g.,][]{Ogura_and_Phillips_1962,Gough1969}---which are appropriate to study convection and mean flows in the interior of low-mass stars where the flow velocities are substantially subsonic and thermal perturbations are small relative to their mean values. Under the anelastic formulation, the fluid equations can be written in the following form:

\begin{gather}
\grad \cdot (\overline{\rho}\bm{u}) = 0\,,\\
\overline{\rho}\left(\dfrac{D\bm{u}}{Dt} + 2\bm{\Omega}_0\cross \bm{u}\right) = -\overline{\rho}\grad\left(\dfrac{P'}{\overline{\rho}}\right) - \dfrac{\overline{\rho}S'}{c_p}\bm{g} + \grad \cdot \bm{\mathcal{D}}\,,\\
\overline{\rho}\overline{T}\dfrac{DS'}{Dt} = \grad \cdot \left(\overline{\rho}\overline{T}\kappa_T\grad S'\right) + Q + \Phi\,.
\end{gather}
In the equations above, $\bm{u}$ is the flow velocity, $\bm{g} = -(GM_{\star}/r^2)\bm{\hat{r}}$ is the acceleration due to gravity (i.e., we ignore the self-gravity of the convection zone, and assume that the entire mass of the star $M_{\star}$ lies below the convection zone), $\bm{\Omega}_0 = \Omega_0\bm{\hat{z}}$ is the rotation vector (aligned with the axis of the spherical coordinate system, $\bm{\hat{z}}$), $\bm{\mathcal{D}}$ is the viscous stress tensor, $Q(r)$ is a function that accounts for radiative heating, and $\Phi$ is the rate of viscous heating. 
Further, $\rho$ is the mass density, $P$ is the pressure, $T$ is the temperature, $S$ is the specific entropy density, and $c_p$ is the specific heat capacity at constant pressure. The thermal diffusivity and the kinematic viscosity are denoted by $\kappa_T$ and $\nu$, respectively.

Note that all thermodynamic quantities are linearized around a temporally steady, spherically symmetric reference state (indicated by an overbar). Fluctuations of these thermodynamic variables about the reference state are denoted by a prime (e.g., the total density is $\rho = \overline{\rho} + \rho'$). We approximate the solar convection zone as a perfect gas, so that the reference state obeys $\overline{P} = \mathcal{R}\overline{\rho}\overline{T}$, where $\mathcal{R}$ is the gas constant. The fluctuations are related by

\begin{equation}
\dfrac{\rho'}{\overline{\rho}} = \dfrac{P'}{\overline{P}} - \dfrac{T'}{\overline{T}} = \dfrac{P'}{\gamma \overline{P}} - \dfrac{S'}{c_p}\, , 
\end{equation}
where $\gamma$ is the adiabatic index.

\subsection{Model Setup}

All simulations in this study use an adiabatically-stratified, polytropic reference state designed to resemble the standard solar model of \citet{ModelS}.  A detailed description of this reference state is provided in \citet{Jones2011} and \citet{Featherstone_and_Hindman_2016}.  We adopt a polytrope of index $n=1.5$ (so that $\gamma = 5/3$). We set $M_{\star} = M_\odot = 1.989\times 10^{33}~\mathrm{g}$, and $c_p = 3.5\times 10^{8}~\mathrm{erg~K^{-1}~g^{-1}}$. 
We adopt an inner radius, $r_\mathrm{inner}$, of 0.718$R_\odot$ and an outer radius, $r_\mathrm{outer}$, of 0.981$R_\odot$ (where $R_{\odot} = 6.959 \times 10^{10}~\mathrm{cm}$ is the solar radius). With this choice, the shell depth is $L \approx 183~\mathrm{Mm}$,  the shell aspect ratio is  $r_{\mathrm{inner}}/r_{\mathrm{outer}} \approx 0.732$ and the mass density varies across the shell by more than two orders of magnitude ($\rho_{\mathrm{inner}}/\rho_{\mathrm{outer}} = e^5 \approx 148$). We set $\kappa_T = \nu = 1.175\times 10^{12}~\mathrm{cm^2~s^{-1}}$.

Following \citet{Featherstone_and_Hindman_2016}, we adopt a heating function of the form

\begin{equation}
Q(r) = C\,\overline{P}(r)\,,
\end{equation}

\noindent and the normalization constant $C$ is chosen such that a solar luminosity $L_{\odot} = 3.846\times 10^{33}~\mathrm{erg~s^{-1}}$ is deposited in the convection zone

\begin{equation}
L_{\odot} = 4\pi\int^{r_{\mathrm{outer}}}_{r_{\mathrm{inner}}} Q(r) r^2 dr\,.
\end{equation}

For all the models in this work, we adopt impenetrable and stress-free boundary conditions for the velocity field. At the inner boundary, we employ a thermally insulating boundary condition ($\partial S'/\partial r|_{r=r_{\mathrm{inner}}} = 0$), and at the outer surface, we enforce that the stellar luminosity $L_{\odot}$ deposited into the convection zone by internal heating is carried out via thermal conduction

\begin{equation}
    \left.\frac{\partial S'}{\partial r}\right|_{r=r_{\mathrm{outer}}} = -\left.\frac{L_{\odot}}{4\pi r^2 \overline{\rho}\overline{T}\kappa_T} \right|_{r=r_{\mathrm{outer}}}\; .
\end{equation}

Finally, the physical variables are represented by truncated expansions on the sphere, with spherical harmonics for the angular directions and Chebyshev polynomials for the radial direction. All the simulations are conducted using the same resolution, where the number of radial, latitudinal, and longitudinal points are $(N_r,\,N_{\theta},\, N_{\phi}) = (256, 1536, 3072)$. After achieving thermal and dynamical equilibration, the models were evolved for at least a thermal diffusion timescale, $L^2/\kappa_T$ ($\sim 10^{3}$--$10^{4}$ rotations in our models).

\subsection{Nondimensional Control Parameters}

Three nondimensional parameters govern the evolution of the flow. These are the flux Rayleigh number $\mathrm{Ra}$, the Ekman number $\mathrm{Ek}$, and the Prandtl number $\mathrm{Pr}$, 
which in this work are defined respectively as

\begin{equation}
\label{eq:dimensionless_numbers}
\mathrm{Ra} \equiv \dfrac{|\Tilde{g}|\Tilde{F}L^4}{\Tilde{\rho}\Tilde{T}c_p \nu\kappa_T^2}, \quad \mathrm{Ek} \equiv \dfrac{\nu}{2\Omega_0 L^2}, \quad \mathrm{Pr} \equiv \dfrac{\nu}{\kappa_T},
\end{equation}
where tildes indicate volume averages over the convective shell. Here, $F$ is the thermal energy flux that derives from the internal heating $Q(r)$ \citep[see][]{Featherstone_and_Hindman_2016}.  

These numbers can be expressed more concisely in terms of timescales \citep[see e.g.,][]{Camisassa_Featherstone_2022}, yielding

\begin{equation}
\label{eq:dimensionless_numbers}
\mathrm{Ra} \equiv \left(\frac{\tau_\nu}{\tau_{\rm ff}}\right)\left(\frac{\tau_\kappa}{\tau_{\rm ff}}\right), \quad \mathrm{Ek} \equiv \left(\frac{\tau_{\Omega}}{\tau_\nu}\right), \quad \mathrm{Pr} \equiv \frac{\tau_\kappa}{\tau_\nu}.
\end{equation}
Here, $\tau_\nu$ and $\tau_\kappa$ are the viscous and thermal diffusion times across the domain respectively, and $\tau_\Omega$ is the rotational timescale.  Separately, $\tau_{\rm ff}$, which is akin to the convective overturning time, is the characteristic time for a cool parcel to freely fall across the convection zone. 

In stellar interiors, these ratios reach extreme values owing to the relative weakness of diffusion when compared to other physical effects.  Under solar conditions, order of magnitude estimates give $\mathrm{Pr} = \mathcal{O}(10^{-6})$, $\mathrm{Ra} = \mathcal{O}(10^{22})$, and $\mathrm{Ek} = \mathcal{O}(10^{-15})$ \citep[e.g.,][]{Schumacher_and_Katepalli_2020}. It is currently impossible, and likely to remain so for decades, to create numerical simulations with solar-like values of the Rayleigh, Ekman, and Prandtl numbers;  the spatio-temporal resolution requirements are simply too high. 

By contrast, the simulations considered here have $\mathrm{Pr} = 1$, $\mathrm{Ra} = 3\times 10^6$, and $\mathrm{Ek} \in [3.6\times 10^{-5}$, \, $2.7\times 10^{-4}$]. Fortunately, direct matching of the nondimensional control parameters is not necessary for fruitful study of large-scale flows. Instead, two conditions must be satisfied to ensure that the models and the Sun are in a similar dynamical regime.  

First, the Rayleigh number (and so too the degree of turbulence) must be sufficiently large that diffusive processes drop out of the leading-order force and thermal balances.  As discussed in \citet{Featherstone_and_Hindman_2016_L,Featherstone_and_Hindman_2016}, this condition is satisfied for the Rayleigh and Ekman numbers under consideration here.
 
The second consideration is that the Coriolis force experienced by convective motions must be sufficiently strong that a solar-like differential rotation, with fast equator and slow poles, is sustained.  When this is not the case, a so-called antisolar differential rotation, with rapidly-rotating poles and a slowly-rotating equator will develop.  The transition between these two regimes is best characterized by the convective Rossby number, Ro$_\mathrm{c}$ defined as

\begin{equation}
\mathrm{Ro_c} = \left(\dfrac{\mathrm{Ra}\,\mathrm{Ek}^2}{\mathrm{Pr}}\right)^{1/2}\ = \frac{\tau_\Omega}{\tau_{\rm ff}} .
\end{equation}
As long as Ro$_\mathrm{c}$ is less than unity, a solar-like differential rotation is sustained \citep[e.g.,][]{Gastine_et_al_2014,Camisassa_Featherstone_2022}.  Using the solar estimates for our nondimensional control parameters noted above, we find that $\mathrm{Ro_c}=\mathcal{O}(10^{-1})$.  This motivates our choice for the range of Ekman numbers explored in this study, which translates to $\mathrm{Ro_c} \in [0.0625, 0.5]$.

\begin{figure}
    \centering
    \includegraphics[width=0.48\textwidth]{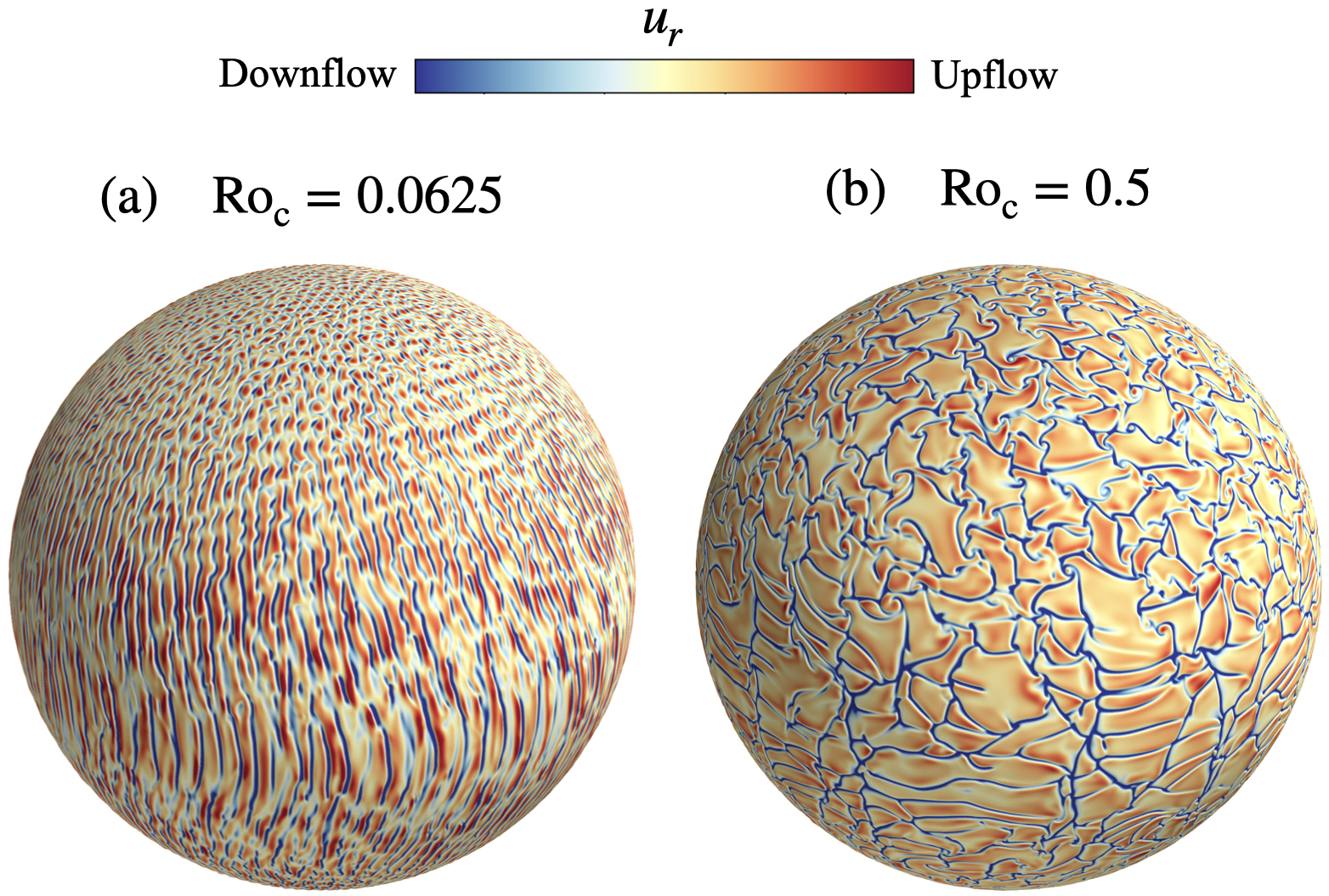}
    \caption{Radial velocity near the upper boundary ($r/r_{\mathrm{outer}} = 0.968$) for the rapidly and slowly rotating cases (panels a and b, respectively). Upflows are indicated in red and downflows in blue. Note that for $\mathrm{Ro_c} = 0.0625$, the flow structure varies across latitude, with vortices dominating the poles, and Taylor columns near the equator. On the contrary, for $\mathrm{Ro_c} = 0.5$, the convective flows are nearly isotropic. Although the Taylor columns are still present, they are less obvious, and can only be seen near the equator as an alignment of the downflow lanes with the rotation axis. Both cases have the same Rayleigh number $\mathrm{Ra}=3\times 10^{6}$ and Prandtl number $\mathrm{Pr}=1$.
    }
    \label{fig:1}
\end{figure}

To illustrate qualitatively the differences in the morphology of the flows for different values of $\mathrm{Ro_c}$, we first present the two extreme cases in our set of models, namely, $\mathrm{Ro_c} = 0.0625$ (strong rotational influence), and $\mathrm{Ro_c} = 0.5$ (moderate rotational influence). Figure~\ref{fig:1} shows snapshots of the radial velocity near the upper surface ($r/r_{\mathrm{outer}}=0.968$) for both cases.  In the low-$\mathrm{Ro_c}$ case (panel a), the flow structures are noteably thinner and more elongated in latitude than those arising in the high-$\mathrm{Ro_c}$ case in panel b.  In that system, only weak north-south alignment of the flow is apparent, and convective structures possess significantly larger horizontal extent.  This variation in convective structure translates into similar variation in the angular momentun transport and, through it, the meridional flow achieved \citep[e.g., ][]{Miesch_2011, Featherstone_2015}.

\subsection{Flow fields} \label{sec:flow_fields}

\begin{figure}
    \centering
    \includegraphics[width=0.48\textwidth]{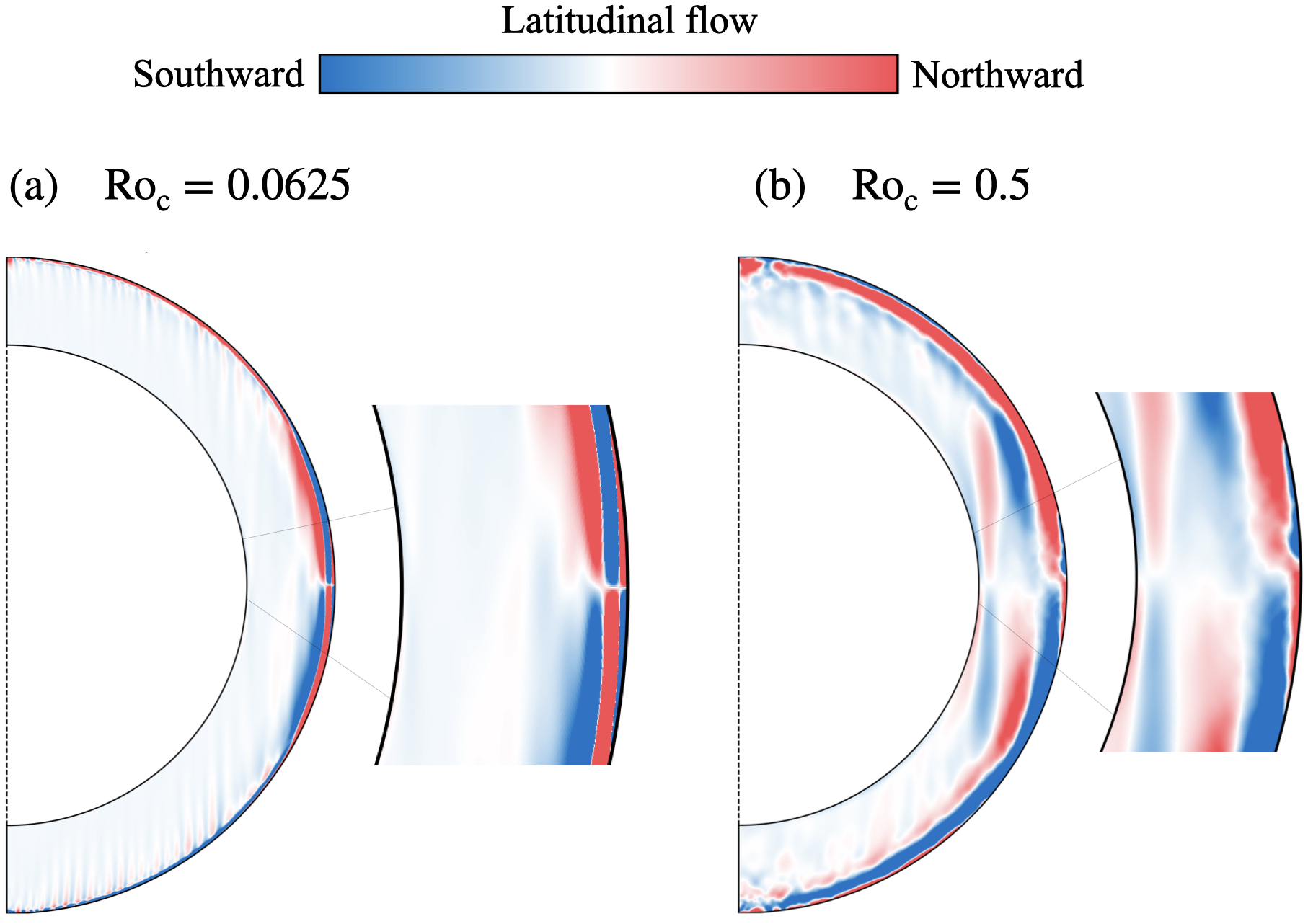}
    \caption{Latitudinal component of the meridional flow, averaged over time and longitude as described in the text. Results are shown for the rapidly and slowly rotating cases (panels a and b, respectively). Northward flows are indicated in red and southward in blue. In both cases, there are multiple cells across the shell depth, although for $\mathrm{Ro_c} = 0.0625$, the cells are confined to the surface. We zoom in on the flows around the equator to better visualize the tiny cells at the surface which occur in a thin Ekman boundary layer that is typically 2 Mm in thickness.}
 \label{fig:2}
\end{figure}

In order to analyze the meridional circulation, we examine the mean flows obtained from the simulations. This is done in two steps. First, we average the flows across the azimuthal ($\phi$) direction. Next, we take a temporal average over the periods when the system is in a statistically stationary state ($\sim 10^3$--$10^4$ rotations). Figure~\ref{fig:2} shows the latitudinal component of the meridional flow, for the same cases discussed previously ($\mathrm{Ro_c} = 0.0625$ in panel a, and $\mathrm{Ro_c} = 1$ in panel b). 

We observe that the flow consists of multiple cells throughout the shell. However, the key difference between the two runs is that for $\mathrm{Ro_c} = 0.0625$, the flows are limited to the surface layers and only at low-latitudes can they penetrate down to approximately half of the convection zone. On the other hand, for $\mathrm{Ro_c} = 0.5$, the cells are present throughout the entire depth of the convection zone, with a spherical shape near the surface and a more cylindrical shape towards the interior. 

Finally, it is worth mentioning that for the range of $\mathrm{Ro_c}$ considered in this work, we have not found any solutions where the meridional circulation is composed of a single cell. Although we do find solutions with single cells for $\mathrm{Ro_c} > 1$, we exclude them deliberately since they have anti-solar differential rotation \citep[e.g.,][]{Gastine_et_al_2014,Camisassa_Featherstone_2022}.

We note that irrespective of the direction of the near-surface meridional flows, which for some of the cases considered here are directed to the poles (resembling the ones in real observations), there exist additional extremely thin circulation cells that hug the outer surface at the equator and at high latitude (a feature that is not present in most observations). These boundary layer cells are confined to the outer $1\%$ of the shell and are the result of the impenetrable boundary condition applied at the upper surface. Such Ekman boundary layers are ubiquitous in rotating systems and should have a thickness proportional to the square root of the Ekman number, Ek. The Sun's Ekman layer is incredibly thin (perhaps 100 m using ${\rm Ek}\sim 10^{-14}$). Our simulations have typical Ekman numbers that are larger than the Sun's, but still small ${\rm Ek}\sim10^{-4}$; hence, the boundary layers should be 1\% of the shell depth, $\sim 2$ Mm. When generating the synthetic observations in Section~\ref{sec:analysis}, we conduct our analyses with and without the boundary layer. While the Ekman layers in our models are significantly larger than the Sun's, they are still sufficiently shallow that the tiny cells do not change our results.

\section{Averaging Kernels} \label{sec:kernels}

\begin{figure}
    \centering
    \includegraphics[width=0.45\textwidth]{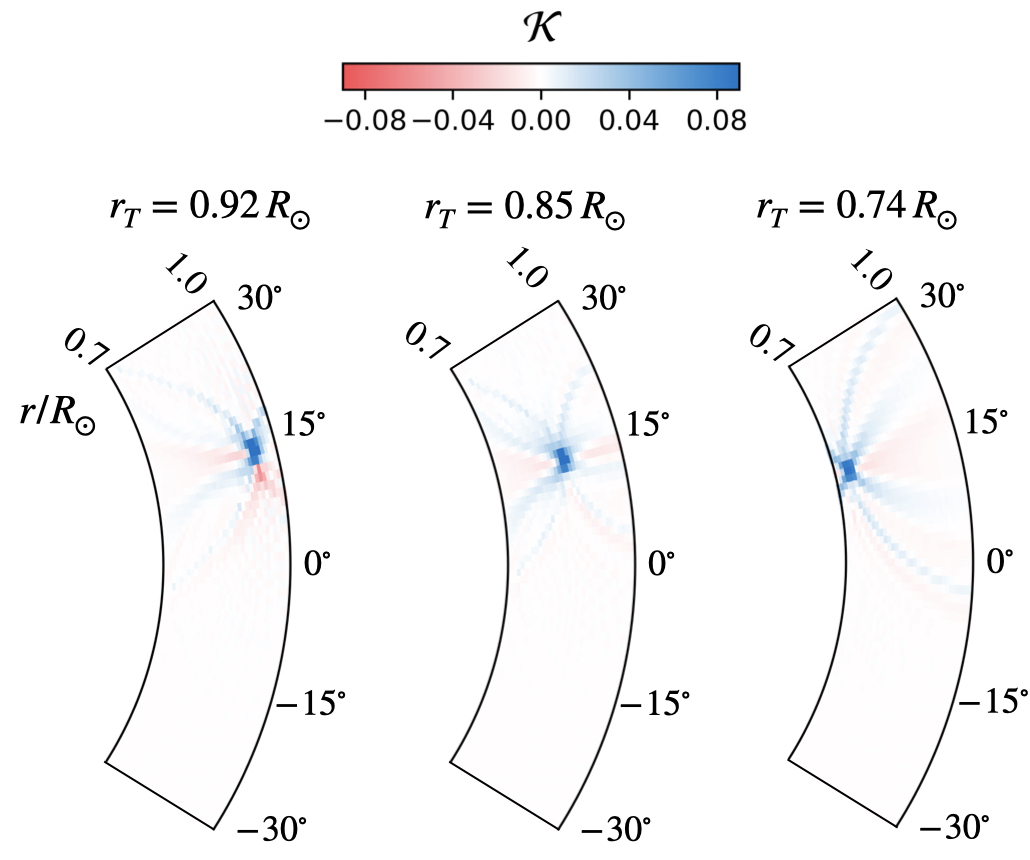}
    \caption{Averaging kernels in \cite{Zhao_2013} for three different target positions, located at latitude $\theta_T = 15^{\circ}$ and at radii $r_T$ = $0.92R_{\odot}$, $0.85R_{\odot}$, and $0.74R_{\odot}$. For clarity, we only show the low-latitude region of the meridional plane, with $\theta \in [-30^{\circ}, + 30^{\circ}]$ and $r/R_{\odot} \in [0.718, 0.996]$. Outside this range, the kernels are essentially zero. The complete set of averaging kernels have information for target depths that lie within the range $ r_T/R_\odot \in [0.637, 0.996]$. Further the kernels are translationally invariant in the target latitude; meaning that the kernels only depend on the difference between the latitude and target latitude, i.e., $\mathcal{K}=\mathcal{K}\left(\theta-\theta_T, r_T, r\right)$. %For a given target location, the kernels have values in the following region of the meridional plane: $\theta \in [-75.6^{\circ},+75.6^{\circ}]$ and $r/R_{\odot} \in [0.536,0.996]$.
    }
    \label{fig:3}
\end{figure}

When using helioseismology to study the meridional circulation in the solar interior, the final product is an estimate of the flow velocity at many positions of interest in latitude and radius (i.e., the target location, $\theta_T$, $r_T$). We make clear that $\theta$ represents latitude and not co-latitude.  The latitudinal component of this flow can be expressed as 

\begin{equation}
u_{\mathrm{obs}}(\theta_T,r_T) = \int_{\mathrm{CZ}} \mathcal{K}(\theta_T,r_T,\theta,r) u_{\mathrm{true}}(\theta,r)dr d\theta\,, \label{eq:u_obs}
\end{equation}
where the integral runs over the spatial extent of the convection zone, $\mathcal{K}$ is the averaging kernel, and $u_{\mathrm{true}}$ is the latitudinal component of the true solar meridional flow. Equation~\eqref{eq:u_obs} can be interpreted as follows: when the flow velocity is measured at a certain location, it represents \emph{a spatial average of the actual velocity}. Ideally, one would desire that the averaging kernels should be as localized as possible, similar to a $\delta$ function. However, regardless of the inversion technique used to obtain the flow field, averaging kernels in deeper regions of the convection zone may not be localized, thus complicating the interpretation of the flow field that is obtained. This can result in smoothing over weak features of small size or contamination from distant locations.

We employ averaging kernels that were originally presented by \cite{Zhao_2013}, who employed time-distance helioseismology \citep[e.g.,][]{Duvall_et_al_1993} to measure the Sun's meridional circulation throughout the convection zone. The kernels were constructed using the ray approximation, which assumes that the acoustic waves are short wavelength and are sensitive only to the flow field along the raypath that connects two points of observation on the solar surface. Further, only sensitivity to the latitudinal component of the flows is considered, which is justified by the fact that the radial component of the flow has a small influence on the travel-time differences. 

 Figure~\ref{fig:3} shows, for several specific target locations, the averaging kernels that we apply later to our simulated meridional flows. These kernels are well localized near the surface, and become broader as the target depth increases, while still remaining localized in the latitudinal direction\footnote{The averaging kernels that we use are for the particular choice of regularization used in the inversions of \cite{Zhao_2013}. Larger or smaller regularization parameters make the kernels less or more localized, respectively.}. We emphasize that the averaging kernels possess negative side lobes that appear both above and below the targeted area. This side-lobe behavior is important because when measuring the flow speed deep in the convection zone, this effect introduces contamination from the fast surface flows (i.e., the integral in Equation~\ref{eq:u_obs} gets contributions from regions that can be far from the location of interest). This is a major point of this paper, and we investigate this effect in Section~\ref{sec:analysis} by applying the averaging kernels to our hydrodynamical simulations.

\begin{figure*}
    \centering
    \includegraphics[width=\textwidth]{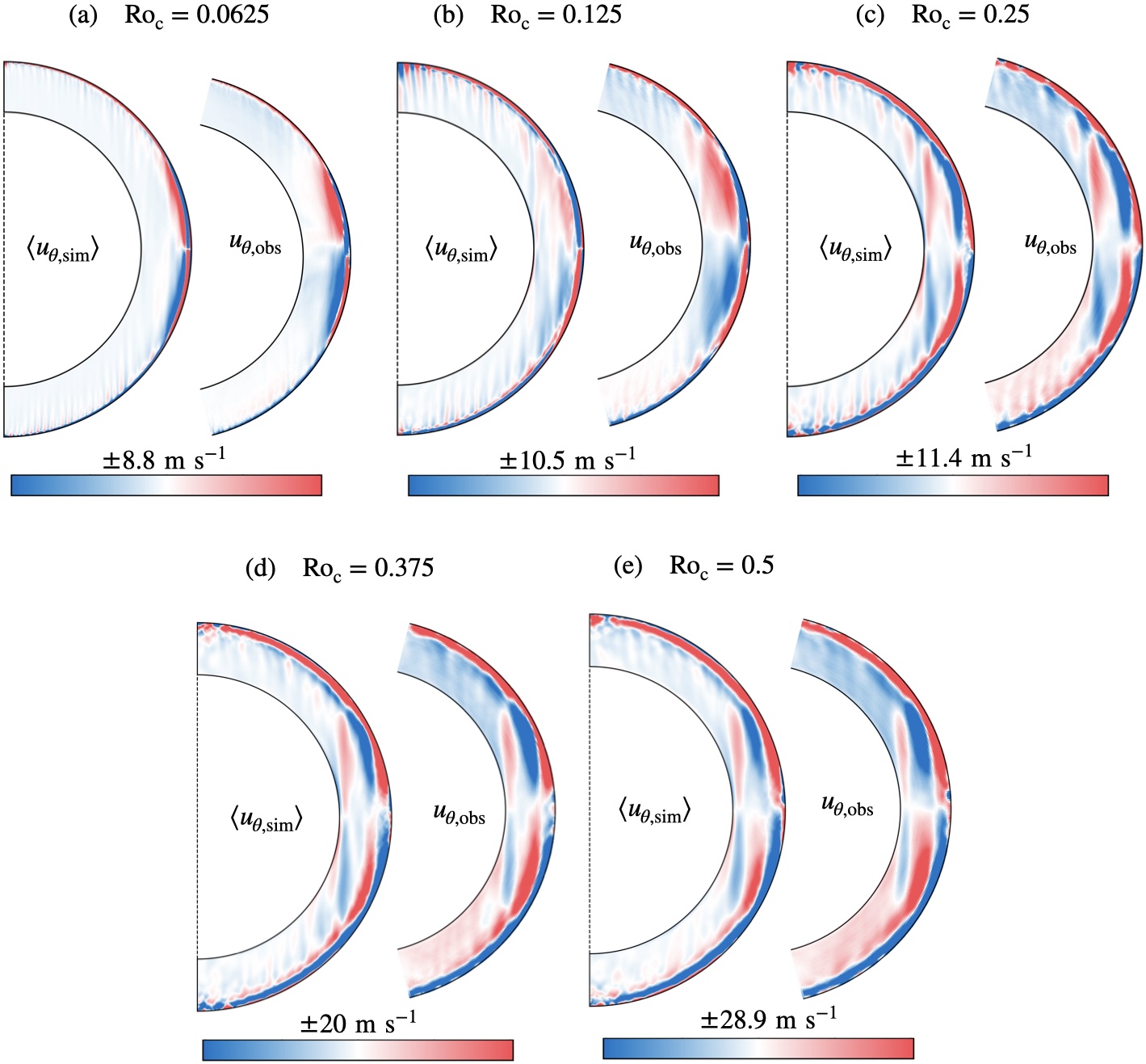}
    \caption{Latitudinal component of the meridional circulation in the simulations, $\langle u_{\theta,\mathrm{sim}}\rangle$, and synthetic observations, $u_{\theta,\mathrm{obs}}$.  Northward flows are indicated in red, and southward in blue. We make clear that $\langle \cdot \rangle$ denotes an average over time and longitude. Different panels show results for runs of different $\mathrm{Ro_c}$. For $\mathrm{Ro_c} = $ 0.0625, 0.125, and 0.25, the fastest flows can reach speeds of $\sim \pm 10~\mathrm{m~s^{-1}}$, while for $\mathrm{Ro_c} =$ 0.375 and 0.5, the fastest velocities are $\sim \pm 20$--$28$ $\mathrm{m~s^{-1}}$. For all the cases presented, there is a good agreement between the simulated and reconstructed flow fields in the surface layers. Differences are more pronounced for deeper flows at mid and high latitudes.
    }
    \label{fig:4}
\end{figure*}
\section{Synthetic Observations}  \label{sec:analysis}

We calculate synthetic ``observed'' flows using Equation~\eqref{eq:u_obs} with the averaging kernels discussed previously, and assuming that $u_{\mathrm{true}}$ is given by the latitudinal component of the mean flows resulting from our simulations, $\langle u_{\theta,\mathrm{sim}}\rangle$ (where $\langle \cdot \rangle$ denotes an average over time and longitude, as explained in Section~\ref{sec:flow_fields}). For each of our simulations, the procedure is as follows: 

\begin{enumerate}
    \item For each target location $(\theta_T,~r_T)$, we make sure that the radial and latitudinal extent of both $\mathcal{K}$ and $\langle u_{\theta,\mathrm{sim}}\rangle$ cover the same range. Then, since the resolution of the simulation data is larger than that of the kernels, we interpolate the simulated flow fields to the same spatial grid as the kernels using a cubic spline interpolation.%\footnote{We also binned the simulation data to the resolution of the kernels, and the results obtained were the same. Nonetheless, we decided to adopt the interpolation method because it is computationally much faster.}. 
    \item Then we compute $\mathcal{K}(\theta_T,r_T,\theta,r)\langle u_{\theta,\mathrm{sim}}\rangle(\theta,r)$, and integrate over all $r$ and $\theta$ as in Equation~(\ref{eq:u_obs}). The result is a scalar that corresponds to an estimate of the flow field at $(\theta_T,r_T)$.

    \item Then, steps 1 and 2 are repeated over the range $r_T/R_{\odot} \in [0.718, 0.981]$, and $\theta_T \in [-75^{\circ}, + 75^{\circ}]$. As a result, a 2D flow field in the meridional plane is generated. This synthetic observation is the final product of applying the averaging kernels to the simulations.
\end{enumerate}

Figure~\ref{fig:4} shows a qualitative comparison between the latitudinal component of the meridional flow from the simulations with the synthetic observations. We find that throughout the domain, small-scale features are broadened and, in many cases, disappear due to the spatial averaging imposed by the width of the averaging kernels. We point the reader to the many narrow high-latitude flow features in panels (c) and (d). On the other hand, large-scale flow patterns, particularly in the upper half of the convection zone, are accurately replicated, even when the flow direction reverses near the surface. However, flows in the lower half of the convection zone, even if they survive being smoothed, can be significantly modified by near-surface flows due to side lobes in the averaging kernels.

\section{Discussion} \label{sec:disc}

\begin{figure*}
    \centering
    \includegraphics[width=\textwidth]{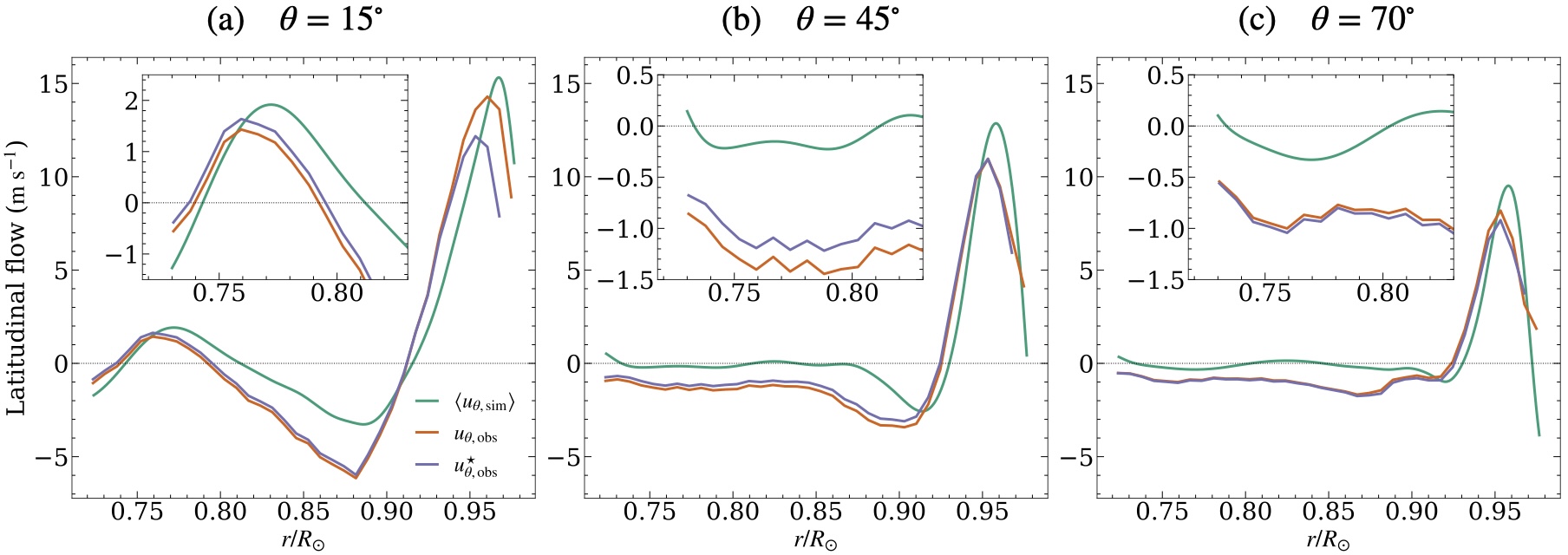}
    \caption{Radial profiles of the flow fields for the simulations and synthetic observations. Results are shown for the case $\mathrm{Ro_c} = 0.5$. To illustrate the minor effect of the thin boundary layer at the surface (confined to the outer $1\%$ of the shell, we show results for synthetic observations including and removing the boundary layer, $u_{\theta,\mathrm{obs}}$ and $u^{\star}_{\theta,\mathrm{obs}}$, respectively)}.
    Panels (a), (b), and (c) show the profiles at $\theta = 15^{\circ}$, $\theta = 45^{\circ}$, and $\theta = 70^{\circ}$, respectively. We select those positions because they exhibit much weaker flows in the deeper layers of the convection zone (as shown in the zoomed-in regions).
    \label{fig:5}
\end{figure*}

We have confirmed that deep, large-scale flow cells can withstand the convolution process if they are sufficiently fast. This begs the question, ``What is the smallest flow speed that can be well-reproduced?". A potential answer can be found when looking into radial profiles of the simulated flows and synthetic observations at different latitudes (see Figure~\ref{fig:5}, which shows such profiles for $\mathrm{Ro_c} = 0.5$, and typifies the results from this analysis for different $\mathrm{Ro_c}$ as well). 

On average, we find that flows whose magnitudes are $\lesssim 1~\mathrm{m~s^{-1}}$ (largely those at mid and high latitudes in our simulations) are hard to recover after the convolution. They get overwhelmed by a spurious flow of $\sim$1 m s$^{-1}$ with the opposite sign. This is clearly seen in the zoomed-in regions of panels b and c. The original flows have speeds of $\pm 0.2~\mathrm{m~s^{-1}}$ and change direction at multiple radial locations. On the contrary, the flows in the synthetic observation have equatorward speeds of roughly $ -1~\mathrm{m~s^{-1}}$, and only change their sign in the surface layers.

Only at low latitudes are multiple cells able to survive the convolution process and show up in the synthetic observations (see Figure~\ref{fig:4} c--d, and Figure~\ref{fig:5} a). There are two reasons for this. First, low-latitude flows are about one order of magnitude faster ($\sim$2 m s$^{-1}$) than flows at the same radii, but higher latitudes. Second, the flows that lie above $0.75 R_\odot$, where the kernels have negative side lobes, switch direction with radius and hence tend to cancel out in the convolution process. 

Finally, as pointed out by \cite{Zhao_2013}, the primary difficulty when measuring weak flows are the random errors that pass through the inversion. Estimates of the random errors at low latitude from \citet{Zhao_2013} (based on two years of data) are typically $\sim$1 m s$^{-1}$ near the surface, and $\sim$5 m s$^{-1}$ in the regions below $0.8R_{\odot}$. We expect that a similar analyses using observations of longer duration would enhance the signal-to-noise ratio throughout the convection zone. For observations spanning a solar cycle, 11 yr, this corresponds to typical errors of $\sim$1.5 m s$^{-1}$. Thus, we find that measurements of the meridional flow averaged over a decade face the unenviable position that the speed and uncertainty of the flow at the base of the convection zone, say 2 \pm 1.5 m s$^{-1}$ \citep[e.g.][]{Zhao_2013, Gizon_2020} are both comparable to the systematic bias caused by the negative side lobes in the averaging kernels. We particularly wish to emphasize that the strong poleward flow at the surface ($\sim$20 m s$^{-1}$) results in a spurious equatorward flow at depth ($\sim$1 m s$^{-1}$), potentially overwhelming any additional cells below, complicating the identification of the return flow. While our findings are directly applicable only to the particular inversion procedure in \cite{Zhao_2013}, they serve as a cautionary note that similar biases may impact other procedures to varying degrees.

\begin{acknowledgements}
We appreciate the commentary by the reviewers, which led to significant improvements in the manuscript. This work was primarily supported by NASA grants 80NSSC18K1125, 80NSSC19K0267 and 80NSSC20K0193.  N.F. and the Rayleigh software were additionally supported by National Science Foundation awards NSF-0949446, NSF-1550901 and NSF-2149126. MC acknowledges grant RYC2021-032721-I, funded by the European Union NextGenerationEU/PRTR and AGAUR/Generalitat de Catalunya grant SGR-386/2021. T.H. was supported by NASA contract NAS5-02139. L.K. acknowledges support from NASA through grant No. 80NSSC21K0455. L.I.M. was supported by NSF award AST-2202253. Computations were conducted with support from the NASA High End Computing (HEC) Program through the NASA Advanced Supercomputing (NAS) Division at Ames Research Center on Pleiades. This work was done in collaboration with the COFFIES DSC which operates under the cooperative agreement 80NSSC22M0162.
\end{acknowledgements}

\bibliography{references}{}

\begin{thebibliography}{}
\expandafter\ifx\csname natexlab\endcsname\relax\def\natexlab#1{#1}\fi

\bibitem[{{Birch} \& {Felder}(2004)}]{Birch2004}
{Birch}, A.~C., \& {Felder}, G. 2004, ApJ, 616, 1261

\bibitem[{{Birch} {et~al.}(2001){Birch}, {Kosovichev}, {Price}, \&
  {Schlottmann}}]{Birch2001}
{Birch}, A.~C., {Kosovichev}, A.~G., {Price}, G.~H., \& {Schlottmann}, R.~B.
  2001, ApJL, 561, L229

\bibitem[{{B{\"o}ning} {et~al.}(2017){B{\"o}ning}, {Roth}, {Jackiewicz}, \&
  {Kholikov}}]{Boning2017}
{B{\"o}ning}, V. G.~A., {Roth}, M., {Jackiewicz}, J., \& {Kholikov}, S. 2017,
  ApJ, 845, 2

\bibitem[{{Camisassa} \& {Featherstone}(2022)}]{Camisassa_Featherstone_2022}
{Camisassa}, M.~E., \& {Featherstone}, N.~A. 2022, ApJ, 938, 65

\bibitem[{{Chen} \& {Zhao}(2017)}]{Chen_and_Zhao_2017}
{Chen}, R., \& {Zhao}, J. 2017, \apj, 849, 144

\bibitem[{{Chen} \& {Zhao}(2018)}]{Chen_and_Zhao_2018}
---. 2018, \apj, 853, 161

\bibitem[{{Christensen-Dalsgaard} {et~al.}(1996){Christensen-Dalsgaard},
  {Dappen}, {Ajukov}, {Anderson}, {Antia}, {Basu}, {Baturin}, {Berthomieu},
  {Chaboyer}, {Chitre}, {Cox}, {Demarque}, {Donatowicz}, {Dziembowski},
  {Gabriel}, {Gough}, {Guenther}, {Guzik}, {Harvey}, {Hill}, {Houdek},
  {Iglesias}, {Kosovichev}, {Leibacher}, {Morel}, {Proffitt}, {Provost},
  {Reiter}, {Rhodes}, {Rogers}, {Roxburgh}, {Thompson}, \& {Ulrich}}]{ModelS}
{Christensen-Dalsgaard}, J., {Dappen}, W., {Ajukov}, S.~V., {et~al.} 1996,
  Science, 272, 1286

\bibitem[{{Duvall} {et~al.}(1993){Duvall}, {Jefferies}, {Harvey}, \&
  {Pomerantz}}]{Duvall_et_al_1993}
{Duvall}, T.~L., J., {Jefferies}, S.~M., {Harvey}, J.~W., \& {Pomerantz}, M.~A.
  1993, Natur, 362, 430

\bibitem[{{Featherstone} {et~al.}(2021){Featherstone}, {Edelmann},
  {Gassmoeller}, {Matilsky}, {Orvedahl}, \& {Wilson}}]{Rayleigh_code_2021}
{Featherstone}, N.~A., {Edelmann}, P. V.~F., {Gassmoeller}, R., {et~al.} 2021,
  {geodynamics/Rayleigh: Rayleigh Version 1.0.0}, Zenodo, v.v1.0.0,  Zenodo,
  doi:10.5281/zenodo.5683601

\bibitem[{{Featherstone} \&
  {Hindman}(2016{\natexlab{a}})}]{Featherstone_and_Hindman_2016_L}
{Featherstone}, N.~A., \& {Hindman}, B.~W. 2016{\natexlab{a}}, ApJL, 830, L15

\bibitem[{{Featherstone} \&
  {Hindman}(2016{\natexlab{b}})}]{Featherstone_and_Hindman_2016}
---. 2016{\natexlab{b}}, \apj, 818, 32

\bibitem[{{Featherstone} \& {Miesch}(2015)}]{Featherstone_2015}
{Featherstone}, N.~A., \& {Miesch}, M.~S. 2015, \apj, 804, 67

\bibitem[{{Gastine} {et~al.}(2014){Gastine}, {Yadav}, {Morin}, {Reiners}, \&
  {Wicht}}]{Gastine_et_al_2014}
{Gastine}, T., {Yadav}, R.~K., {Morin}, J., {Reiners}, A., \& {Wicht}, J. 2014,
  MNRAS, 438, L76

\bibitem[{{Giles}(2000)}]{Giles_2000}
{Giles}, P.~M. 2000, PhD thesis, Stanford University, California

\bibitem[{{Gizon} {et~al.}(2020){Gizon}, {Cameron}, {Pourabdian}, {Liang},
  {Fournier}, {Birch}, \& {Hanson}}]{Gizon_2020}
{Gizon}, L., {Cameron}, R.~H., {Pourabdian}, M., {et~al.} 2020, Sci, 368, 1469

\bibitem[{{Glatzmaier}(1984)}]{Glatzmaier_1984}
{Glatzmaier}, G.~A. 1984, JCoPh, 55, 461

\bibitem[{{Gough}(1969)}]{Gough1969}
{Gough}, D.~O. 1969, JAtS, 26, 448

\bibitem[{{Greer} {et~al.}(2013){Greer}, {Hindman}, \& {Toomre}}]{Greer_2013}
{Greer}, B., {Hindman}, B., \& {Toomre}, J. 2013, in Astronomical Society of
  the Pacific Conference Series, Vol. 478, Fifty Years of Seismology of the Sun
  and Stars, ed. K.~{Jain}, S.~C. {Tripathy}, F.~{Hill}, J.~W. {Leibacher}, \&
  A.~A. {Pevtsov}, 199

\bibitem[{{Hanasoge}(2022)}]{Hanasoge_2022}
{Hanasoge}, S.~M. 2022, LRSP, 19, 3

\bibitem[{{Hartlep} {et~al.}(2013){Hartlep}, {Zhao}, {Kosovichev}, \&
  {Mansour}}]{Hartlep_2013}
{Hartlep}, T., {Zhao}, J., {Kosovichev}, A.~G., \& {Mansour}, N.~N. 2013, \apj,
  762, 132

\bibitem[{{Hathaway}(2012)}]{Hathaway_2012}
{Hathaway}, D.~H. 2012, \apj, 760, 84

\bibitem[{{Hathaway} {et~al.}(2022){Hathaway}, {Upton}, \&
  {Mahajan}}]{Hathaway2022}
{Hathaway}, D.~H., {Upton}, L.~A., \& {Mahajan}, S.~S. 2022, Frontiers in
  Astronomy and Space Sciences, 9, 419

\bibitem[{{Jackiewicz} {et~al.}(2015){Jackiewicz}, {Serebryanskiy}, \&
  {Kholikov}}]{Jackiewicz_et_al_2015}
{Jackiewicz}, J., {Serebryanskiy}, A., \& {Kholikov}, S. 2015, ApJ, 805, 133

\bibitem[{{Jiang} {et~al.}(2010){Jiang}, {I{\c{s}}ik}, {Cameron}, {Schmitt}, \&
  {Sch{\"u}ssler}}]{Jiang_2010}
{Jiang}, J., {I{\c{s}}ik}, E., {Cameron}, R.~H., {Schmitt}, D., \&
  {Sch{\"u}ssler}, M. 2010, \apj, 717, 597

\bibitem[{{Jones} {et~al.}(2011){Jones}, {Boronski}, {Brun}, {Glatzmaier},
  {Gastine}, {Miesch}, \& {Wicht}}]{Jones2011}
{Jones}, C.~A., {Boronski}, P., {Brun}, A.~S., {et~al.} 2011, Icar, 216, 120

\bibitem[{{Komm} {et~al.}(1993){Komm}, {Howard}, \& {Harvey}}]{Komm_1993}
{Komm}, R.~W., {Howard}, R.~F., \& {Harvey}, J.~W. 1993, SoPh, 147, 207

\bibitem[{{Lin} \& {Chou}(2018)}]{Lin_Chou_2018}
{Lin}, C.-H., \& {Chou}, D.-Y. 2018, \apj, 860, 48

\bibitem[{{Mandal} {et~al.}(2017){Mandal}, {Bhattacharya}, {Halder}, \&
  {Hanasoge}}]{Mandal_2017}
{Mandal}, K., {Bhattacharya}, J., {Halder}, S., \& {Hanasoge}, S.~M. 2017,
  \apj, 842, 89

\bibitem[{{Mandal} {et~al.}(2018){Mandal}, {Hanasoge}, {Rajaguru}, \&
  {Antia}}]{Mandal2018}
{Mandal}, K., {Hanasoge}, S.~M., {Rajaguru}, S.~P., \& {Antia}, H.~M. 2018,
  \apj, 863, 39

\bibitem[{{Miesch}(2005)}]{Miesch_2005}
{Miesch}, M.~S. 2005, LRSP, 2, 1

\bibitem[{{Miesch} \& {Hindman}(2011)}]{Miesch_2011}
{Miesch}, M.~S., \& {Hindman}, B.~W. 2011, \apj, 743, 79

\bibitem[{{Ogura} \& {Phillips}(1962)}]{Ogura_and_Phillips_1962}
{Ogura}, Y., \& {Phillips}, N.~A. 1962, JAtS, 19, 173

\bibitem[{{Rajaguru} \& {Antia}(2015)}]{Rajaguru_2015}
{Rajaguru}, S.~P., \& {Antia}, H.~M. 2015, \apj, 813, 114

\bibitem[{{Rajaguru} \& {Antia}(2020)}]{Rajaguru_2020}
{Rajaguru}, S.~P., \& {Antia}, H.~M. 2020, in Astrophysics and Space Science
  Proceedings, Vol.~57, Dynamics of the Sun and Stars; Honoring the Life and
  Work of Michael J. Thompson, ed. M.~J.~P.~F.~G. {Monteiro}, R.~A.
  {Garc{\'\i}a}, J.~{Christensen-Dalsgaard}, \& S.~W. {McIntosh}, 107--113

\bibitem[{{Schad} {et~al.}(2013){Schad}, {Timmer}, \& {Roth}}]{Schad2013}
{Schad}, A., {Timmer}, J., \& {Roth}, M. 2013, ApJL, 778, L38

\bibitem[{{Schumacher} \& {Sreenivasan}(2020)}]{Schumacher_and_Katepalli_2020}
{Schumacher}, J., \& {Sreenivasan}, K.~R. 2020, RvMP, 92, 041001

\bibitem[{{Stejko} {et~al.}(2022){Stejko}, {Kosovichev}, {Featherstone},
  {Guerrero}, {Hindman}, {Matilsky}, \& {Warnecke}}]{Stejko_2023}
{Stejko}, A.~M., {Kosovichev}, A.~G., {Featherstone}, N.~A., {et~al.} 2022,
  \apj, 934, 161

\bibitem[{{Ulrich}(2010)}]{Ulrich_2010}
{Ulrich}, R.~K. 2010, \apj, 725, 658

\bibitem[{{Upton} \& {Hathaway}(2014{\natexlab{a}})}]{Upton_2014b}
{Upton}, L., \& {Hathaway}, D.~H. 2014{\natexlab{a}}, \apj, 792, 142

\bibitem[{{Upton} \& {Hathaway}(2014{\natexlab{b}})}]{Upton_2014a}
---. 2014{\natexlab{b}}, \apj, 780, 5

\bibitem[{{{\v{S}}vanda} {et~al.}(2007){{\v{S}}vanda}, {Kosovichev}, \&
  {Zhao}}]{Svanda_2007}
{{\v{S}}vanda}, M., {Kosovichev}, A.~G., \& {Zhao}, J. 2007, \apjl, 670, L69

\bibitem[{{Wang} {et~al.}(1989){Wang}, {Nash}, \& {Sheeley}}]{Wang_89}
{Wang}, Y.~M., {Nash}, A.~G., \& {Sheeley}, N.~R., J. 1989, Sci, 245, 712

\bibitem[{{Zhao} {et~al.}(2013){Zhao}, {Bogart}, {Kosovichev}, {Duvall}, \&
  {Hartlep}}]{Zhao_2013}
{Zhao}, J., {Bogart}, R.~S., {Kosovichev}, A.~G., {Duvall}, T.~L., J., \&
  {Hartlep}, T. 2013, \apjl, 774, L29

\bibitem[{{Zhao} {et~al.}(2012){Zhao}, {Nagashima}, {Bogart}, {Kosovichev}, \&
  {Duvall}}]{Zhao_2012}
{Zhao}, J., {Nagashima}, K., {Bogart}, R.~S., {Kosovichev}, A.~G., \& {Duvall},
  T.~L., J. 2012, \apjl, 749, L5

\bibitem[{{Zhao} {et~al.}(2016){Zhao}, {Stejko}, \& {Chen}}]{Zhao_et_al_2016}
{Zhao}, J., {Stejko}, A., \& {Chen}, R. 2016, SoPh, 291, 731

\end{thebibliography}
\bibliographystyle{aasjournal}

\end{document}